# Link Quality and MAC-Overhead aware Predictive Preemptive Routing Protocol for Mobile Ad hoc Network

Moussa ALI CHERIF[1] and Mohamed Kamel FERAOUN[1]

[1]Computer Science Department, Sidi Bel Abbes University
Sidi Bel Abbes, Algeria

**Abstract**

In Ad Hoc networks, route failure may occur due to less received power, mobility, congestion and node failures. Many approaches have been proposed in literature to solve this problem, where a node predicts pre-emptively the route failure that occurs with the less received power. However, this approach encounters some difficulties, especially in scenario without mobility where route failures may arise. In this paper, we propose an improvement of AODV protocol called LO-PPAODV (Link Quality and MAC-Overhead aware Predictive Preemptive AODV). This protocol is based on new metric combine more routing metrics (Link Quality, MAC Overhead) between each node and one hop neighbor. Also we propose a cross-layer networking mechanism to distinguish between both situations, failures due to congestion or mobility, and consequently avoiding unnecessary route repair process. The LO-PPAODV was implemented using NS-2. The simulation results show that our approach improves the overall performance of the network. It reduces the average end to end delay, the routing overhead, MAC errors and route errors, and increases the packet delivery fraction of the network.

**Keywords:** *Ad-Hoc networks, AODV, PPAODV, QoS, Cross layer*

## 1. Introduction

An ad hoc network consists of mobile nodes, which communicate with each other through multi-hop routes. Nodes cooperate with their neighbors to route data packets to their final destinations. In ad hoc networks, network topology is changing continuously because of the node movement. To maintain the communication between nodes, many routing protocols have been proposed, which are classified under two categories: table-driven and on-demand routing protocols.

On-demand routing protocols discover routes only when the source needs to send packets. Therefore, there is almost no route maintenance overhead, whereas the route discovery before data transmission increases the delay. However, if the link failure happened, nodes should inform the sources to change the existing route and retransmit the packets that were lost due to link failure. Therefore, on-demand routing protocols increase delay and decrease the successful packet arrival ratio. This causes the reduction of the packet delivery ratio.

Several approaches have been proposed [3,4] to flexibly anticipate link failure by adding a function that predicts the link failure in one of the popular on-demand routing protocols which is Ad hoc On-demand Distance Vector (AODV).

Previous approaches encounter some difficulties, especially in scenario without mobility. The problem is that these approaches predict link failures based of RSS information and interpret that it happened due to node mobility, where actually it was due to congestion. Therefore, the process of route repair should not be performed since it increases even more the congestion, decreasing the overall performance of the network.

Transmitting information to a neighboring node in MAC layer is preceded by the exchange of Request To Send (RTS)/Clear To Send (CTS) frames. If this communication fails, the MAC layer waits (back off time) and retries later. After several failed attempts, the MAC layer informs the routing layer using a cross layer interaction. In our approach, the cause of that unsuccessful communication is sent to the routing layer. If the last received power of the destination node indicates that it is reachable, the routing layer is informed, using the variable xmit_reason with the value XMIT_REASON_HIGH_RSS. Depending on this information a node will decide whether it performs a route repair or not.

In this paper, we propose Link Quality and MAC-Overhead aware Predictive Preemptive Ad hoc On-Demand Distance Vector (LO-PPAODV), it is an on-demand routing protocol based on new metric combine more routing metrics (Link Quality, MAC Overhead), that aims to create congestion-free routes by making use of information gathered from the MAC layer. Also we propose a cross-layer networking mechanism to distinguish between both situations, failures due to congestion or mobility, and consequently avoiding unnecessary route repair process, where we use a "Route Failure Prediction Technique" based on the Lagrange interpolation for estimating whether an active link is about to fail or will fail.

The rest of the paper is organized as follows. Section 2 describes related works; section 3 describes an overview of AODV; the proposed protocol is presented in section 4 and its performance is evaluated and compared with that of PPAODV in section 5. Some conclusions are given in section 6.

## 2. Related Works

In [8] Norman and Joseph propose an energy efficient routing protocol (HLAODV) for heterogeneous sensor networks with the goal of finding the nearest base station or sink node. Hence the problem of routing is reduced to finding the nearest base station problem in heterogeneous networks.

Xiaoqin, Jones and Jayalath In [10] have proposed the Congestion Aware Routing protocol for Mobile ad hoc networks (CARM). Also they have proposed a congestion-aware routing metric which was employed data-rate, MAC overhead, and buffer queuing delay.

In [11] the authors have proposed a link availability-based QoS-aware (LABQ) routing protocol for mobile ad hoc sensor networks based on mobility prediction and link quality measurement, in addition to energy consumption estimate was proposed.

In [12] Yi and Shakkottai have developed a fair hop-by-hop congestion control algorithm with the MAC constraint was being imposed in the form of a channel access time constraint, using an optimization-based framework. They have used a Lyapunov-function-based approach.

Chen and Heinzelman [13] have proposed a QoS-aware routing protocol that were an admission control scheme and a feedback scheme to meet the QoS requirements of real-time applications was incorporated.

Chenxi and Corson [14] have developed a QoS routing protocol for ad hoc networks using TDMA. They aims to establish bandwidth guaranteed QoS routes in small networks whose topologies were changed at low to medium rate.

In [15] CRP, a congestion-adaptive routing protocol for MANETs, was proposed by Tran and Raghavendra. CRP tried to prevent congestion from occurring in the first place, rather than dealing with it reactively.
In [16] a cross-layer designs among physical, medium access control and routing (network) layers, using Received Signal Strength (RSS) was proposed by Chandran and Shanmugavel. Their object was energy conservation, unidirectional link rejection and reliable route formation in mobile ad hoc networks.

Xia, Ren and Liang [17] have introduced a method for cross-layer design in mobile ad hoc networks. They have used fuzzy logic system (FLS) to coordinate physical layer, data link layer and application layer for cross-layer design.

Authors in [19] have proposed a link availability-based QoS-aware (LABQ) routing protocol for mobile ad hoc networks based on mobility prediction and link quality measurement, in addition to energy consumption estimate.

Baboo and Narasimhan [20] have proposed a hop-by-hop congestion aware routing protocol which employs a combined weight value as a routing metric, based on the data rate, queuing delay, link quality and MAC overhead. Among the discovered routes, the route with minimum cost index is selected, which is based on the node weight of all the in-network nodes.

In [21] Bisengar, Rziza and Ouadou have proposed an improvement of AODV protocol called AMAODV (Adaptative Mobility aware AODV). This protocol is based on new metric combine more routing metrics (distance, relative velocity, queue length and hop count) between each node and one hop neighbor.

In [23] a model was proposed that extends the existing AODV routing protocol to accommodate additional QoS constraints for minimum session bandwidth required for an application. Extensions are added to the current AODV messages during route discovery which specify the QoS requirements.

In [24] Sedrati, Bilami and Benmohamed propose a new variant based on the AODV which gives better results than the original AODV protocol with respect of a set of QoS parameters and under different constraints, taking into account the limited resources of mobile environments (bandwidth, energy). The proposed variant (M-AODV) suggests that the discovering operation for paths reconstruction should be done from the source. It also defines a new mechanism for determining multiple disjoint (separated) routes.

In order to reduce the number of broken routes, the authors propose [25] a novel reliable routing algorithm using fuzzy applicability to increase the reliability during the routing selection. In the proposed algorithm source chooses a stable path for nodes mobility by considering nodes position/ velocity information. Also they propose novel method for rout maintenance, in this protocol before breaking packet transmitted path a new one is established.

### 2.1 Link failure prediction methods

In [3], a Predictive Preemptive AODV (PPAODV) was proposed which predicts the link failure using the Received Signal Strength (RSS) has been proposed. The prediction method uses Lagrange interpolation, which approximates the process of RSS by means of n-dimensional function with information of past RSS.

PPAODV [3] discovers a new route before the active route becomes obsolete and changes the route smoothly by predicting a RSS of data packets at the Predict Time $t_{PT}$ from the past information of RSS. PPAODV [3] sets Discovery Period $T_{DP}$ as the minimum time that a node can exchange one data with the neighboring node.

In [4], the authors have proposed a High Precision - PPAODV (HPPPAODV) which is an amelioration of PPAODV. HPPPAODV can improve the prediction accuracy ratio by 1) using the Newton interpolation, 2) adding the chance of acquisition of RSS to reduce the error margin of RSS that is affected by the influence of the thermal noise and fading and 3) predicting the value of the Discovery Period $T_{DP}$ by the number of hop in a route.

## 3. AODV Overview

AODV [1,2] is an on-demand routing protocol. Route discovery is initiated only when a source node needs to communicate with a destination for which it does not have a route in its routing table. To discover a route to a destination, the source node broadcasts a route request message (RREQ) that contains a request ID. If a node receives a RREQ that it has already received, it drops the request. Otherwise, it stores the address of the node from which it received the request so as to establish a reverse route to the source that it uses later. If the RREQ reaches a node that has a route to the destination, the node sends, over the reverse route, a route reply message (RREP) to the source. The reply message contains the number of hops needed to reach the destination from the node. If the RREQ reaches the destination, it sends a route reply to the source over the reverse route.

Intermediate nodes that do not have a path to the destination re-broadcast the request. As the RREP is sent back to the source over the reverse path each node stores the address of the node that sent the reply. The forward path thus determined from the source to the destination is used for sending packets to their destination. AODV uses sequence numbers maintained for the different destinations so to guarantee freshness of routing information.

A link breaks when a node within an active route moves out of the transmission range of its upstream neighbor. When a link break occurs, the node upstream the break invalidates, in its routing table, all routes become unusable due to the loss of the link. It then creates a Route Error (RERR) message, in which it lists the destinations that have become unreachable because of the loss of the link. The RERR is sent to all source nodes that use the link. This procedure is named global repair. AODV also includes a local repair mechanism to locally recover from link losses. Local repair is triggered when a link break occurs between nodes within an active route. In this repair, the node upstream the break tries to find alternative sub-paths to the destinations of packets that it has received, but is unable to forward them (packets) because of the link break.

## 4. The proposed LO-PPAODV

4.1 Protocol Overview

4.1.1 Link Quality Estimation

In this paper, two-ray ground model is adopted.
This model [27] considers both the direct path and a ground reflection path. The model gives more accurate prediction at a long distance than the free space model. The received power is predicted by:

$$P_r(d) = \frac{P_t G_t G_r h_t^2 h_r^2}{d^4 L} \qquad (1)$$

Where *Pt* is the transmitted signal power.
*Gt* and *Gr* are the antenna gains of the transmitter and the receiver respectively.
*L* is the system loss, *d* is the distance between transmitter and receiver. *ht* and *hr* are the heights of the transmit and receive antennas respectively.
In this paper, we suppose that the transmit range of each node is equivalent.

So, the link quality $\quad Lq = Pr \qquad (2)$

4.1.2 Estimating MAC Overhead

We consider IEEE 802.11 MAC with the distributed coordination function (DCF). It has the packet sequence as request-to-send (RTS), clear-to-send (CTS), and data, acknowledge (ACK). The short inter frame space (SIFS) is the amount of time between the receipt of one packet and the transmission of the next. Then the channel occupation due to MAC contention is:

$$Coc = t_{RTS} + t_{CTS} + 3t_{SIFS} \qquad (3)$$

Where $t_{RTS}$ and $t_{CTS}$ are the time consumed on RTS and CTS, respectively and $t_{SIFS}$ is the SIFS period.
Then the MAC overhead $OH_{MAC}$ can be represented as:

$$OH_{MAC} = C_{oc} + t_{ac} \qquad (4)$$

Where $t_{ac}$ is the time taken due to access contention.
The amount of MAC overhead is mainly dependent upon the medium access contention, and the number of packet collisions. That is, $OH_{MAC}$ is strongly related to the congestion around a given node.

$OH_{MAC}$ can become relatively large if congestion is incurred and not controlled, and it can dramatically decrease the capacity of a congested link.

LO-PPAODV employs a combined weight metric in its cost function. The node weight metric *fpd* which assigns a cost to each link in the network. Weight function *fpd* combines the link quality *Lq* and MAC overhead $OH_{MAC}$ to select optimal paths.

The *fpd* for the link from node *i* to a particular neighboring node is given by :

$fpd = (Lq)/(OH_{MAC})$ (5)

LO-PPAODV is reactive routing protocol; no permanent routes are stored in nodes. The source node initiates route discovery procedure by broadcasting. The RREQ message is organized as detailed in Table 1.

Table 1. RREQ message in LO-PPAODV

| .TYPE u_int8_t | Reserved | HOP COUNT u_int8_t |
|---|---|---|
| RREQ BROADCAST ID     u_int32_t ||| 
| DESTINATION IP ADDRESS     nsaddr_t |||
| DESTINATION SEQUENCE NUMBER     u_int32_t |||
| SOURCE IP ADDRESS     nsaddr_t |||
| SOURCE SEQUENCE NUMBER     u_int32_t |||
| Cost *fpd*     double |||

When the source node issues a new RREQ (figure1), the *fpd* value in RREQ is initialized to 65536 ($2^{16}$).
After the destination node receives the first RREQ, it starts to wait for a period of time to receive enough RREQs. Then it selects the route with the biggest cost *fpd* value and sends back a Route Reply (RREP) to the source node via the selected route.
If there are multiple routes with the same *cost* the route with the smallest hop count is selected. Let $p_c$ be the chosen path and $p_a$ the set of all possible paths. Then the chosen path fulfills:

$$cost(p_c) = \min \max_{p_j \in p^a}\left(cost(p_j)\right)$$

Upon receiving the RREP, an intermediate node records the previous hop and relays the packet to the next hop.
If a node detects a link break during route maintenance phase, it sends a Route Error (RERR) packet to the source node. Upon receiving the RERR, the source node initiates a new round of route discovery.

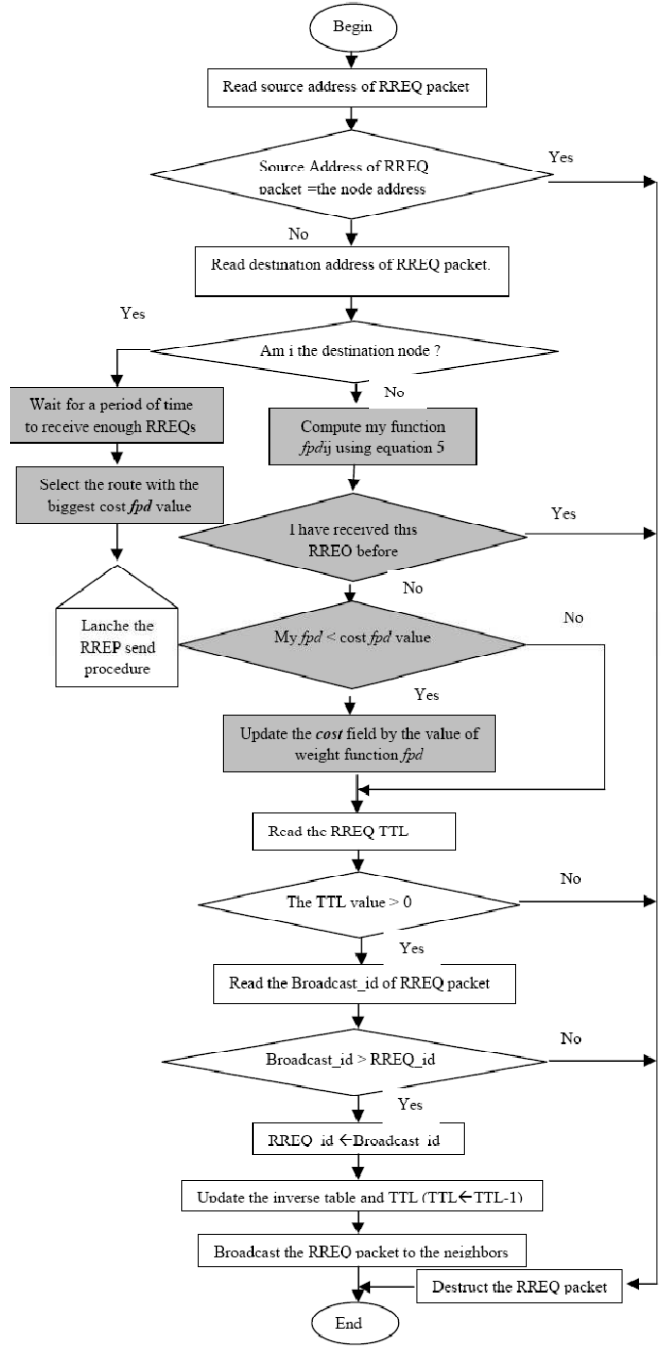

Fig. 1 flow char for RREQ in LO-PPAODV

### 4.2 The proposed mechanism for Congestion Control

In LO-PPAODV we implemented a cross layer approach that tracks the RSS of received data packet from each neighboring node in order to know when an adjacent node is near enough for a successful transmission.

We use a "Route Failure Prediction Technique" based on the Lagrange interpolation (6) for estimating whether an active link is about to fail or will fail, and it can distinguish between both situations; link error at MAC layer was due to congestion and due to mobility of nodes to avoid the unnecessary route repair process. The Predict Time ($t_{PT}$) is calculated as (7) and the Discovery Period $T_{DP}$ can be calculated as (8).

$$P(t_{PT}) = \left(\frac{(tPT-t2)(tPT-t3)}{(t1-t2)(t1-t3)} \times P1\right) + \left(\frac{(tPT-t1)(tPT-t3)}{(t2-t1)(t2-t3)} \times P2\right) + \left(\frac{(tPT-t1)(tPT-t2)}{(t3-t1)(t3-t2)} \times P3\right) \quad (6)$$

Where $P(t_{PT})$ is the value of RSS at $t_{PT}$. P1 –P3 and t1–t3 are 1st –3rd RSS and their received time respectively.

$$t_{PT} = t3 + T_{DP} \quad (7)$$

$$T_{DP} = T_{warning} \times n_{A\text{-}S} + T_{RREQ} \times n_{S\text{-}D} + T_{RREP} \times n_{S\text{-}D} \quad (8)$$

Where, $T_{warning}$, $T_{RREQ}$ and $T_{RREP}$ represent the transmission time of warning packet, RREQ packet and RREP packet, respectively. Also $n_{A–S}$ and $n_{S–D}$ represent the number of hops between node "A" to node "S" of the active route and number of hops between node S to node D of a new route, respectively.

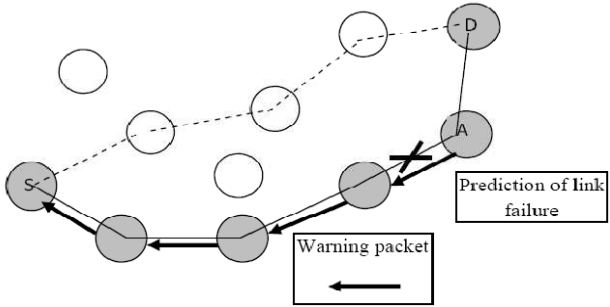

Fig. 2 Node A predicts link failure

### 4.2.1 Extension of MAC layer

AODV interprets a link failure (in MAC layer) as a broken link, even when it was caused by congestion at the receiver. The sender node should know why communication was impossible. We implemented an approach that tracks the RSS of received data packet from each neighboring node in order to know when an adjacent node is near enough for a successful transmission. If lost packets were due to congestion and high traffic, AODV triggers route repair, and this can affect the network performance. If lost packets is due to low signal quality or misrouted packets, then route repair is needed because the receiver is not reachable.

Afterward, the signal strength of neighboring nodes can be used to detect the reason for lost packets, distinguishing between congestion and broken links due to mobility, because in the last case, the receiver is unreachable and its signal strength is now available. The implementation is divided into two parts; the first part keeps the last three received signals from a node in an array, and computes RSS using Lagrange Interpolation (from the received data packets) if the signal is weak enough and the node moving away, the MAC layer sends a Request To Send (RTS) and the second part decides the kind of message (link failure, either due to errors or due to congestion using signal strength of neighboring nodes) to be sent to the upper layer, whenever the communication is impossible but the destination node is in the transmission range of the sender.

Transmitting information to a neighboring node in MAC layer is preceded by the exchange of Request To Send (RTS)/Clear To Send (CTS) frames. If this communication fails, the MAC layer waits (back off time) and retransmits later. After several unsuccessful attempts, the MAC layer informs to the routing layer that communication failed. In our approach, the reason for that unsuccessful communication is sent to the routing layer. If the last received power (the result of Lagrange interpolation) of the destination node indicates that it is reachable, the routing layer is informed, using the variable xmit_reason with the value XMIT_REASON_HIGH_RSS. In this case, the routing layer should interpret that communication to destination was impossible, not because of a broken link but rather congestion, therefore, route maintenance is not needed. If that is not the reason delivered to the routing layer, a route maintenance process is required.

### 4.2.2 Extension of AODV

When a node tries to communicate with a neighboring node and this communication failed (after several attempts, MAC layer sends an error to the routing layer). AODV [1] interprets that the neighboring node is not present anymore and communication failure was due to mobility.

In a scenario without mobility communication failures may arise, but AODV will interpret that it was due to mobility, where actually, it was due to congestion. Therefore, the process of route repair should not be performed since it increases even more the congestion, decreasing the overall performance of the network. The proposed amelioration will make AODV [1] capable to distinguish between both situations, avoiding the route repair process when the link error at MAC layer was due to congestion and not due to mobility of nodes. In our approach, when a node is not able to communicate with a neighboring node, MAC layer informs to the upper layer that there was a problem including whether the neighboring node is still reachable or not (see figure 3). Therefore, the sender node does not perform route maintenance if it was informed that the neighboring node is still reachable.

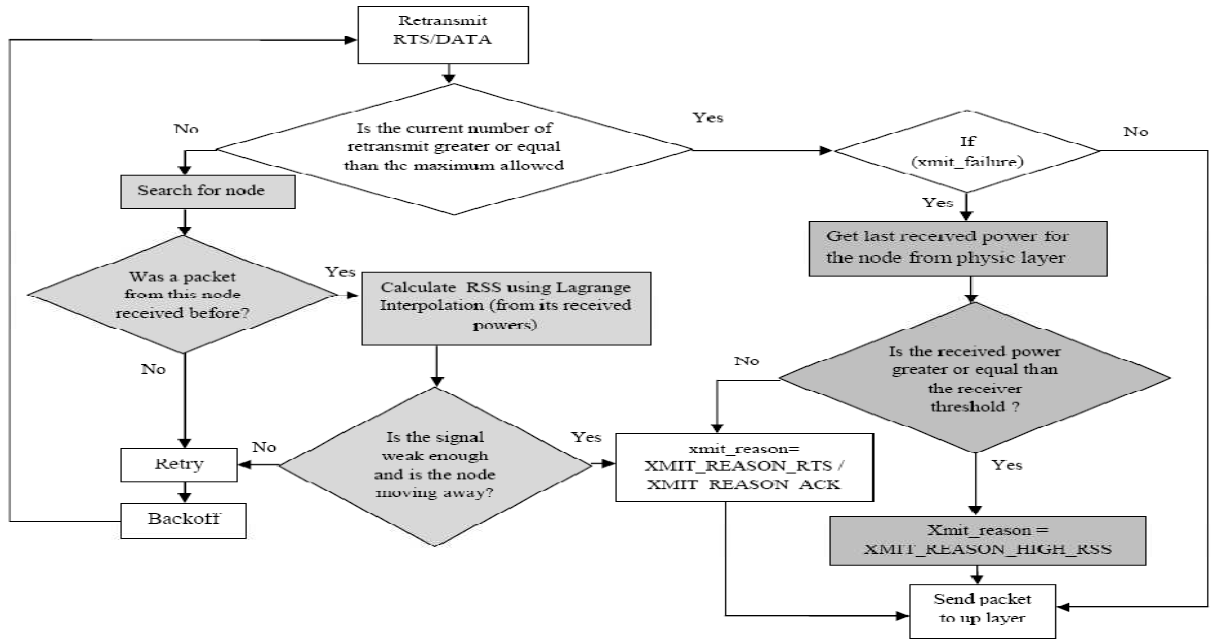

Fig. 3 The proposed approach that uses the Lagrange interpolation is shown here, this diagram shows also how MAC layer informs to the routing layer, when several attempts to communicate to the receiver node failed [28].

## 5. Simulation and Performance Results

We have used the implementation of AODV [1] in the NS simulator version 3.35 [5]. Our results are based on the simulation of 50 wireless nodes forming an ad hoc network moving about in an area of 1500 meters by 300 meters for 200 seconds of simulated time. The physical radio characteristics of each mobile node's network interface, such as the antenna gain, transmission power, and receiver sensitivity, are chosen to approximate the Lucent WaveLAN [6] direct sequence spread spectrum radio.

The movement scenario files used for each simulation are characterized by a pause time. Each node begins the simulation by selecting a random destination in the simulation area and moving to that destination at a speed distributed uniformly between 0 and 10 meters per second. It then remains stationary for pause time seconds. This scenario is repeated for the duration of the simulation. We carry out simulations with movement patterns generated for 5 different pause times: 0, 20, 40, 80 and 200 seconds. A pause time of 0 seconds corresponds to continuous motion, and a pause time of 200 (the length of the simulation) corresponds to limited motion. Constant bit rate (CBR) sources are used in the simulations. The packet rate is 4 packets/sec when 10, 20, 30 and 40 sources are assumed. The performance metrics used to evaluate performance are:

*Packet delivery ratio*: The ratio of the data packets delivered to the destination to those generated by the CBR sources. This should be maximized.

*Average end-to-end delay of data packets:* This includes all possible delays caused by buffering during route discovery, queuing at the interface queue, retransmission delays at the MAC layer, and propagation and transfer times. This should be minimized.

*Normalized routing load*: The number of routing packets transmitted per data packet delivered to the destination. This should be minimized.

*Route errors*: Each time PPAODV or LO-PPAODV performs a route error process at sender; it is registered and showed in the graphic (figure 18). A route error in PPAODV triggers a route maintenance process provoking more control traffic in the network. Usually these kinds of errors are due to broken links because of the mobility of nodes, but they may arise from collision of packets, as well. These errors should be minimized.

We report the results of the simulation experiments for the Predictive Preemptive AODV protocol (PPAODV) and for LO-PPAODV.

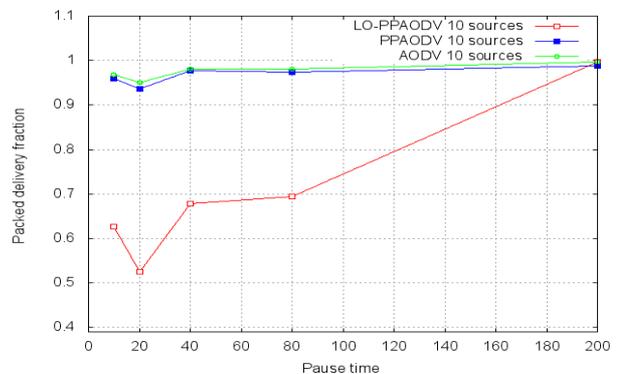

Fig. 4 Packet delivery fraction 10 sources

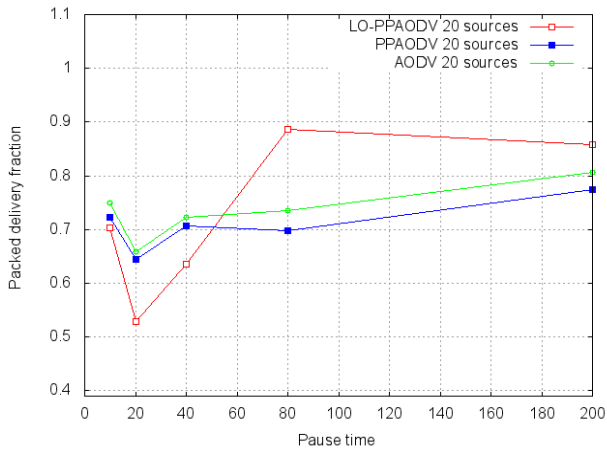

Fig. 5 Packet delivery fraction 20 sources

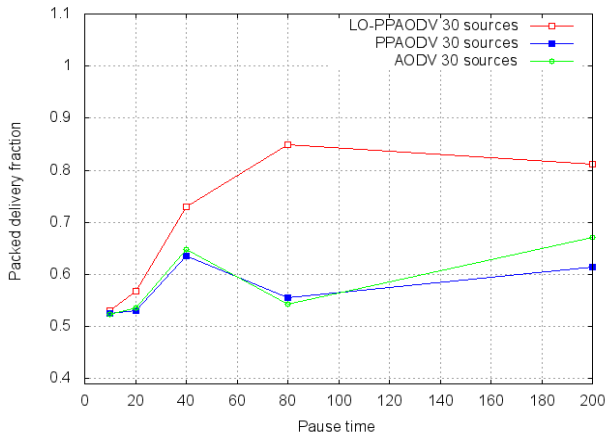

Fig. 6 Packet delivery fraction 30 sources

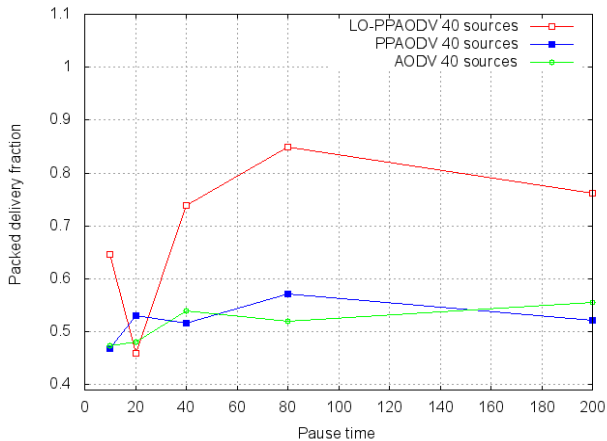

Fig. 7 Packet delivery fraction 40 sources

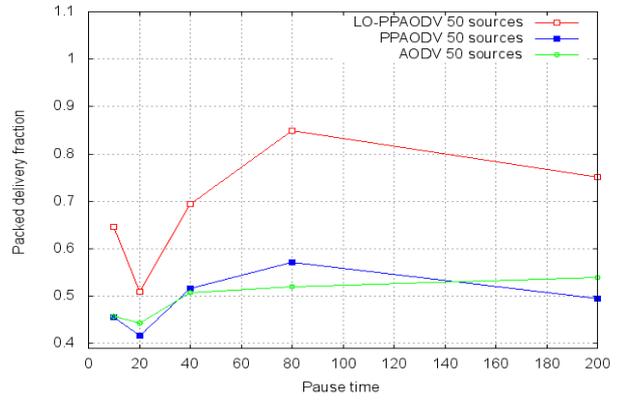

Fig. 8 Packet delivery fraction 50 sources

Figure 4, figure 5, figure 6, figure 7 and figure 8 represents the simulation results for the delivery ratio metric of 10 sources, 20 sources, 30 sources, 40 sources and 50 sources respectively. It can be seen that the method proposed can result in significant performance gains. (The results show that LO-PPAODV outperforms PPAODV significantly when the number of sources increases 30 sources, 40 sources and 50 sources see figure 5 to figure8). We observe that the packed delivery fraction increases significantly when the number of sources increases.

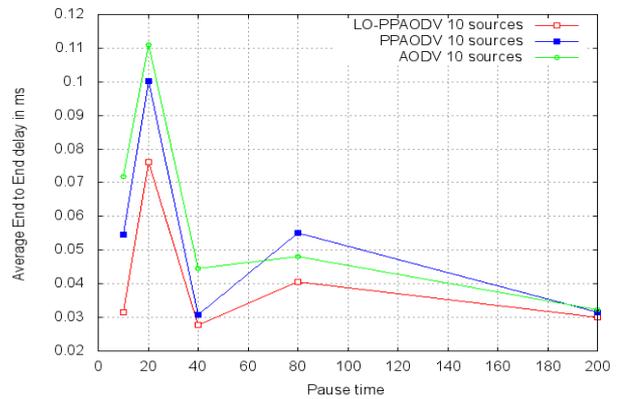

Fig. 9 Average End to end delay 10 sources

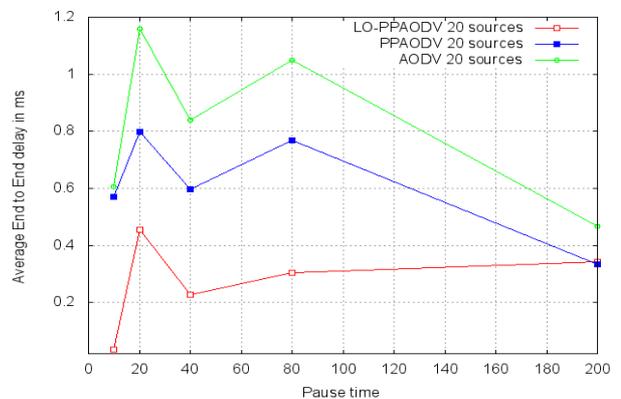

Fig. 10 Average End to end delay 20 sources

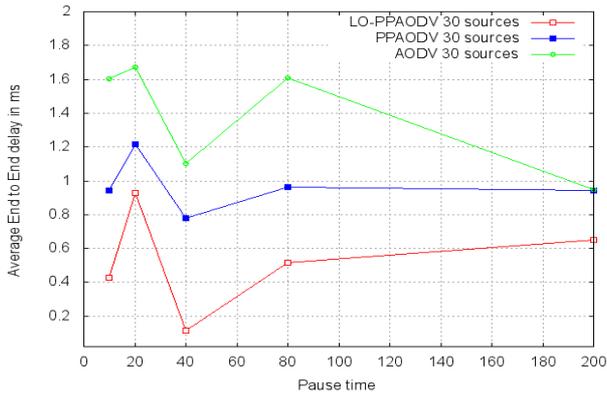

Fig. 11 Average End to end delay 30 sources

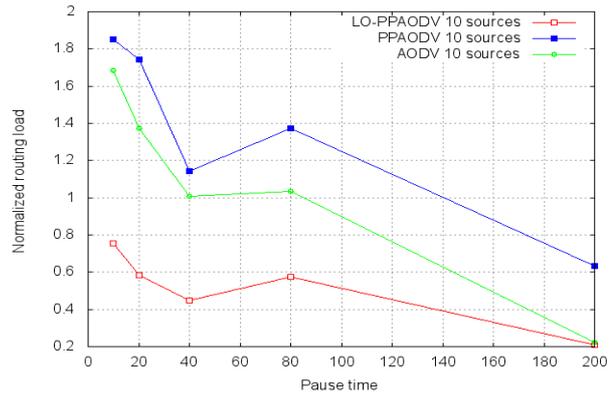

Fig. 14 Normalized routing load 10 sources

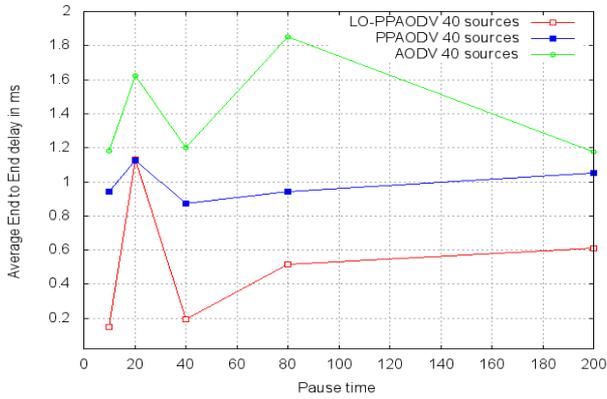

Fig. 12 Average End to end delay 40 source

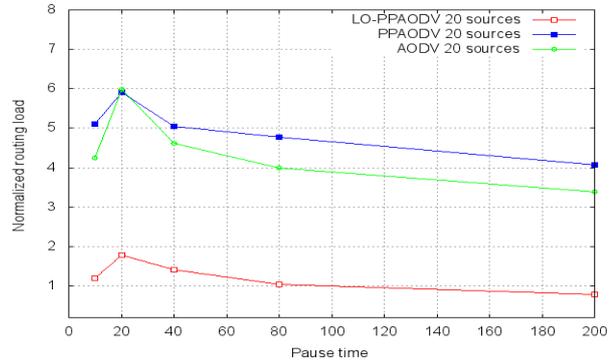

Fig. 15 Normalized routing load 20 sources

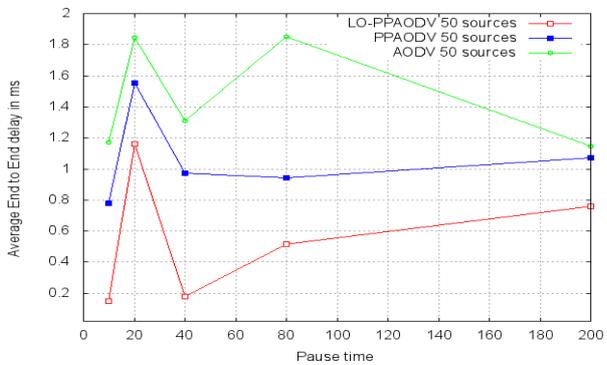

Fig. 13 Average End to end delay 50 sources

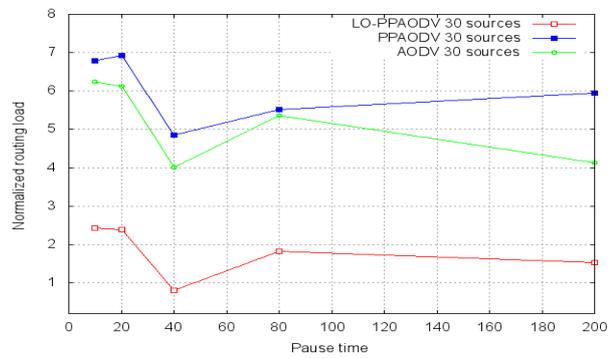

Fig. 16 Normalized routing load 30 sources

In figure 9, figure 10, figure 11, figure 12 and figure 13 the results obtained for the end-to-end delay metric of 10 sources, 20 sources, 30 sources, 40 sources and 50 sources respectively are presented. We observe that the end-to-end delay increases significantly when the number of sources increases. The delay is affected by the route repair procedure because data packets are buffered until an alternative route is found. The results show that LO-PPAODV outperforms PPAODV significantly when the number of sources increase and the motion is low. Figure 12 shows a gain of about 90% of LO-PPAODV over PPAODV, for 40 sources in the pause time 200s.

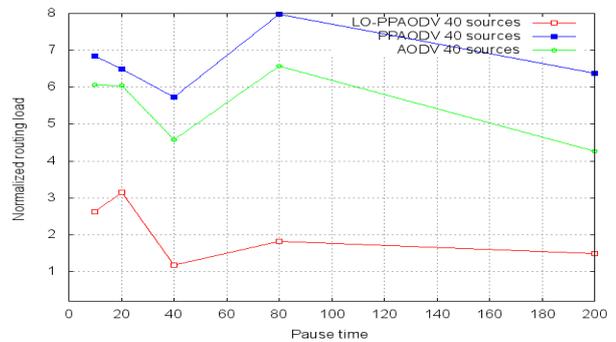

Fig. 17 Normalized routing load 40 sources

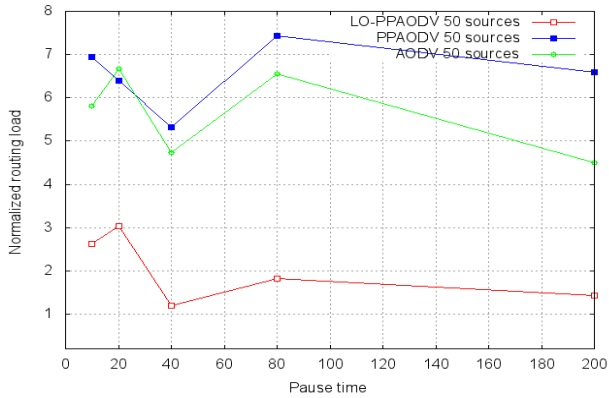

Fig. 18 Normalized routing load 50 sources

Figure 14, figure 15, figure 16, figure 17 and figure 18 show how mobility and number of sources affect the communication overhead. The overhead is high when node motion is high; this is due to the fact that it is difficult to obtain an alternative link to replace a broken one when motion is high. It is also observed that the overhead is low when the number of sources is low. This results from the fact that many sources may share one or more paths, which decreases the communication overhead. It can be observed from figure 16 that the biggest gains of LO-PPAODV over PPAODV is of 300% less and happen with 80s of pause time and 40 sources. This has a good impact on energy because the number of control packets generated is low.

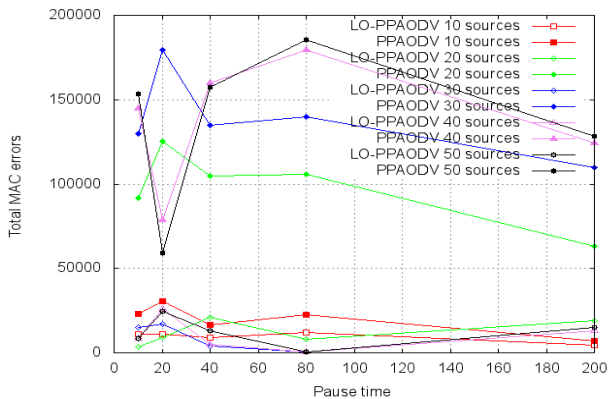

Fig. 19 Total MAC errors

In Figure 19 we observe in the vertical axis the dropped packets from the MAC layer for PPAODV and LO-PPAODV. There were found 4 different types of MAC errors: collision, retry exceed count, MAC busy or duplicate packet. These errors should be minimized. MAC errors are increasing using PPAODV (figure 19), because of the mobility of nodes (from a minimum of 0m/s to a maximum of 10m/s). This mobility, as well, causes a high number of route errors (figure 20) and therefore more routing overhead and packet loss. We see from Figure 19 that LO-PPAODV outperforms PPAODV significantly; MAC errors are decreased for all sources (up to 600% errors less for 50 sources).

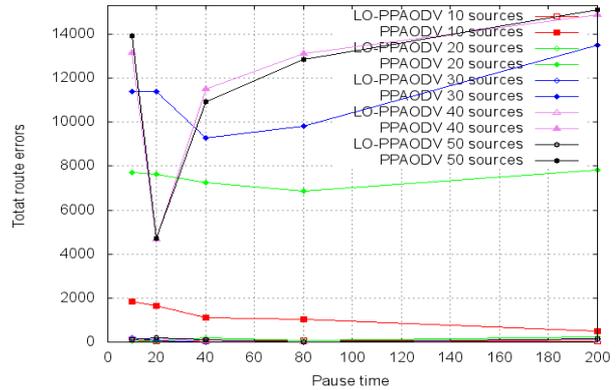

Fig. 20 Total route errors

As conclusion, it is worthwhile to point out that the graphics are related between themselves, since less MAC errors and less route errors provokes lower normalized routing load in the network. As normalized routing load is decreasing, the nodes are able to transmit more data packets; therefore, a higher packed delivery fraction is obtained (up to 300% for 50 sources and happen with 200s of pause time see figure 8).

## 6. Conclusions

In this paper, we have proposed a Link Quality and MAC-Overhead aware Predictive Preemptive Ad hoc On-Demand Distance Vector (LO-PPAODV). There are two main contributions in this work. One is the protocol is based on new metric combine more routing metrics (Link Quality, MAC Overhead) another is the proposition of a cross-layer networking mechanism to distinguish between both situations, failures due to congestion or mobility; by the usage of the "Route Failure Prediction Technique" based on the Lagrange interpolation for estimating whether an active link is about to fail or will fail.

Simulation results show that the average and to end delay of LO-PPAODV is less than that of PPAODV. Also normalized routing load of LO-PPAODV is smaller than that of PPAODV. It can be noticed from this study that the packet delivery fraction is more than that of PPAODV especially when the number of sources is superior to 10 sources. We can see also that MAC errors and route errors are increasing using LO-PPAODV.